# Higher-order topological phase in an acoustic fractal lattice

Junkai Li[1#], Qingyang Mo[1#], Jian-Hua Jiang[2] and Zhaoju Yang[1*]

[1]Department of Physics, Interdisciplinary Center for Quantum Information and Zhejiang Province Key Laboratory of Quantum Technology and Device, Zhejiang University, Hangzhou, China

[2]School of Physical Science and Technology, and Collaborative Innovation Center of Suzhou Nano Science and Technology, Soochow University, Suzhou, China

#These authors contributed equally to this work

*Email: zhaojuyang@zju.edu.cn.

**Abstract:**

Higher-order topological insulators, which support lower-dimensional topological boundary states than the first-order topological insulators, have been intensely investigated in the integer-dimensional systems. Here, we provide a new paradigm by presenting experimentally a higher-order topological phase in a fractal-dimensional system. Through applying the Benalcazar, Bernevig, and Hughes model into a Sierpinski carpet fractal lattice, we uncover a squeezed higher-order phase diagram featuring the abundant corner states, which consist of zero-dimensional outer corner states and 1.89-dimensional inner corner states. As a result, the codimension is now 1.89 and our model can be classified into the fractional-order topological insulators. The non-zero fractional charges at the outer/inner corners indicate that all corner states in the fractal system are indeed topologically nontrivial. Finally, in a fabricated acoustic fractal lattice, we experimentally observe the outer/inner corner states with local acoustic measurements. Our work demonstrates a higher-order topological phase in an acoustic fractal lattice and may pave the way to the fractional-order topological insulators.

Higher-order topological insulators (HOTIs) are a newly developed topological phase which hosts higher-codimensional topological states localized at the "boundaries of boundaries" [1–7] with lower dimensionality. There are various methods to induce a higher-order topological phase. A quintessential example is the quadrupole insulator based on the Benalcazar, Bernevig, and Hughes (BBH) model [1], which has vanishing dipole polarization but a quantized quadrupole moment. Due to the extended higher-order bulk-boundary correspondence principle, a two-dimensional (2D) quadrupole insulator supports 0D in-gap corner states and 1D gapped edge states. Another canonical approach is to directly generalize the 1D SSH model to the higher-dimensions [8–11]. For example, by changing the relative strength of the intra- and inter-hopping in a 2D system, we can induce a second-order topological phase whose topological invariants are characterized by the dipole polarization, instead of the quadrupole moment. These approaches have enriched the research on HOTIs as well as the conventional topological insulators [12,13] over the past decade, given the feasibility of designing artificial crystals where the coupling scheme can be easily engineered. So far, HOTIs have been extensively studied in the classical-wave systems, such as in photonics [14–17], acoustics [18–22] and electric circuits [23–25] and have led to new frontiers in interactions with non-Hermiticity [26,27], disorder [28,29] and nonlinearity [30,31].

On the other hand, a series of intriguing advances have been made in the realization of topological states in the fractals [32–35], which are characterized by their self-similarity and fractional dimensions [36]. Intuitively, fractal lattice will turn off the topological protection [37] due to the lack of a well-defined bulk and the subsequent bulk-edge correspondence. However, a recent report showed that Floquet topological fractal insulators can be realized in a photonic lattice of Sierpinski gasket consisting of helical optical waveguides [35]. The upcoming experimental observation [38] will be the first example of the first-order topological insulators in the fractals. More recently, a preprint suggested that higher-order topological phases can be found in the quantum fractals featuring outer and inner corner modes [39,40], which bridged the gap between the fractality and HOTIs. However, up to now, the experimental verification has not been completed.

In this work, we present an experimental realization of a HOTI in an acoustic fractal lattice. Our model applies the BBH model in a 1.89-dimensional Sierpinski carpet composed of the sonic resonators. First, we calculate the phase diagram based on the real-space quantum many-body multipole operators [41,42] and unveil that the fractality squeezes the nontrivial region of the quadrupole phase by a factor of 0.35. In the topological phase region, the eigenvalues as a function of intra-couplings are in the form of a fractal 'butterfly' and include abundant corner states at the outer and different inner corners. The calculated results of fractional charges [43–45] indicate that both the outer and inner corner states are topologically nontrivial. In experiments, we design the above lattice model by utilizing sonic resonators [46,47] and measure the local responses of the system. By spanning over all the lattice sites, we experimentally observe the abundant topological corner states at the outer and inner corners with the dimensions corresponding to 0 and 1.89 respectively. The fractional dimension of the corner states has never been predicated and therefore makes our work significantly distinctive from the conventional topological insulators. Our work therefore demonstrates an experimental realization of the HOTI in an acoustic fractal lattice.

Our lattice model of the Sierpinski carpet is shown in Fig. 1(a). Through the box-counting method, the dimension of the fractal lattice is $d_f = ln8/ln3 = 1.893$ [see Supplementary Information (SI), section 1]. The minimum primitive cell is constructed by the unit cell of the BBH model, which is depicted in Fig.1(b). The thin and thick lines represent the intra-hopping $t_1$ and the inter-hopping $t_2$ and the red (green) lines indicate positive (negative) couplings, which as a result can introduce an effective magnetic flux $\phi = \pi$ for each plaquette. The details of the specific Hamiltonian describing our fractal model are shown in SI, section 2.

With the elaborately designed fractal model, we calculate the energy spectrum as a function of the intra-hopping $t_1$ and present the result in Fig. 1(c, d). The inter-hopping $t_2$ hereafter is set to be unity for simplicity. The grey and blue curves represent the 'bulk' (interior of the lattice) and edge states, which are similar to those in a HOTI in the integer-dimensional systems [see SI, section 3]. Surprisingly, as can be seen in the center dashed box in panel (c), a large number of corner states intersect at the center of the energy spectrum ($t_1 = 0$ and $E = 0$) and are in the form of a fractal 'butterfly', which is fundamentally different from the traditional HOTI based on BBH model [see SI, section 3]. The red lines indicate the outer corner states that are degenerate within the range from $t_1 = -0.35$ to $t_1 = 0.35$, which is self-consistent with the calculated result of the quadrupole moment, as shown in Fig. 1(e). The yellow and green curves represent two different types of inner corner states A and B.

To determine the topological phase diagram of the fractal quadrupole insulator, we need to resort to the real-space quantized quadrupole moment [41,42] because of the fact that the fractal lattice lacks translational symmetry and Bloch's theorem fails. The formula can be written as:

$$Q_{xy} = \frac{1}{2\pi} Im \log \left[ \det(U^\dagger \hat{Q} U) \sqrt{\det(\hat{Q}^\dagger)} \right] \quad mod \ 1, \tag{1}$$

where $\hat{Q} = exp[i2\pi\hat{x}\hat{y}/(L_x L_y)]$ is a diagonal unitary matric, $\hat{x}$ ($\hat{y}$) is the position operator along the horizontal (vertical) direction, $L_x$ ($L_y$) is the length of lattice in $x$ ($y$) directions and $U$ is a matrix whose columns are the eigenvectors of the Hamiltonian under the condition of half-filling. The quadrupole moments are quantized to 0 or 1/2, which indicates a trivial or higher-order topological phase. In Fig. 1(e), we plot the quadrupole moment $Q_{xy}$ as a function of intra-hopping $t_1$ for the fractal (red dots) and square (grey dots) lattice respectively. In the fractal case, $Q_{xy}$ jumps from 0 to 0.5 at $t_1 = \pm 0.35$, and remains 0.5 within the range of $-0.35 < t_1 < 0.35$, which indicates a squeezed topological region compared to that of the 2D lattice shown as the grey dotted curve. Note that in panel (d,e), we mark the topological phase regions by blue shading.

Having presented the fractal 'butterfly' and topological phase diagram, we move forward on the abundant states carried by the fractal model. we plot the energy spectrum with fixed $t_1/t_2 = 0.13$ in Fig. 2(a), and the enlarged view of all corner states in the inset. First, we focus on four outer corner states (red dots), which are degenerate at the energy of zero. The sum distribution of these four states is localized at four outer corners of the fractal lattice, as shown in Fig. 2(b), which is the key feature of the higher-order topological phase. To verify the squeezed topological phase, we set $t_1/t_2 = 0.8$, which corresponds to the nontrivial phase in the 2D lattice but trivial phase in the fractal system, and find no topological corner states [see SI, section 4]. Second, as can be seen in Fig. 2(c, d), there exist a large number of inner corner states labeled A and B, corresponding to yellow and cyan dots shown in panel (a), localized at the corners of the inner voids with different sizes. Note that, different from the outer corners spanning over an angle of $\pi/2$, the inner corners in the fractal lattice span over $3\pi/2$, which consequently leads to the equal split of the inner corner state into two edge sites next to the corner of the inner void. In addition, when $t_1 = 0$, an additional type of inner corner modes based on the trimers will appear exactly at the energy of the 'bulk' states based on the tetramers [see SI, section 5]. However, with a finite strength of intra-hopping, these inner corner modes are mixed with the 'bulk' states and cannot be distinguished [see SI, section 6]. Note that as the number of generations increases, lots of self-similar inner corner states emerge, which form a more complex fractal 'butterfly' [see SI, section 7], and the topological phase diagram stays squeezed [see SI, section 8]. Furthermore, in Fig. 2(e), we show the sum distribution of the edge states corresponding to the blue dots depicted in Fig. 2(a).

In light of the above discussion, we have found corner states in both exterior and interior of the fractal lattice. It raises a question: are these two types of corner states both topologically nontrivial? To answer this question, we calculate the fractional charges [43–45], which are the fractional parts of the spectral charges [see SI, section 9]. As we can see in Fig. 2(f), the fractional charge at the outer corner equals to 0.5, which is an important characteristic of the higher-order quadrupole insulators. Interestingly, the two corner sites of each inner corner have a fractional charge of 0.25, which together contribute to a fractional charge of 0.5 per inner corner. The reason of the counter-intuitive result is that the inner corners spanning over $3\pi/2$ can split the charge of 0.5 into two halves. More details of the spectral charge are shown in SI, section 9. Since there are non-zero fractional charges at both the inner and outer corners, we conclude here that all corner states are topologically nontrivial and robust against considerable disorder [see SI, section 10].

With the intriguing states presented in the fractal-dimensional model, we are now interested in investigating the dimensionality of the states. Through the box-counting method [48], we calculate the dimension of lattice, corner states and edge states by the formula:

$$D = \lim_{n \to \infty} \frac{\ln(N)}{\ln(N_l)}, \qquad (2)$$

where $N$ is the number of lattice sites or topological states and $N_l$ is the number of length sites on one side. In Fig. 2(g), we plot the ratio $lnN/lnN_l$ for the inner corner, outer corner, edge, and full lattice as a function of different generations G(n) [see SI, section 1]. We can see that as the number of generations increases, the dimension of the outer corners converges to 0 (green dots). Combined with the 1.89D lattice (gray dots), we arrive at the codimension of 1.89 and get a fractional-order topological insulator. Surprisingly, the dimensions of inner corner and edge states are both 1.89, which corresponds to the codimension of 0. These counter-intuitive results have never been reported before and indicate that the HOTIs in the fractal model are significantly different from the conventional HOTIs.

To experimentally demonstrate the higher-order topological phase in a fractal system, we design and fabricate the acoustic fractal lattice consisting of sonic resonators with direct 3D printing. The photo of the real structure in the experiment is shown in Fig. 3(a). More details can be found in SI, section 11. Each cuboid resonator represents a lattice site in the theoretical model. The cross-section of the tube connecting nearest-neighboring resonators corresponds to the strength of coupling, which in the following is fixed with the ratio of $t_1/t_2 = 0.13$ [see SI, section 12] for further demonstration. In experiments, a speaker generating sound waves is inserted into the sample through a square pinhole at one side of a resonator. At the same time, a microphone inserted into another circular pinhole at the other side of the same resonator collects the response data, which can be considered as the local density of states [LDOS, see SI section 13 for more details]. This measurement spans over all the resonators of the sample in order to obtain the field distributions [see SI, section 11].

In Fig. 3(b), we show the measured intensity spectra for four resonators at the outer corner, inner corner, edge, and 'bulk' sites labeled as '1', '2', '3', and '4' in panel (a). The 'bulk' response spectrum (green curve) has two branches centered at the frequency of 4872 Hz and 5674 Hz, which indicates the existence of a wide frequency gap between the two peaks. The intensity of the outer corner response (red curve) is mainly distributed in the gap, and has a single peak located at the frequency of 5382 Hz. As a comparison, the intensity of the inner corner response (yellow curve) is distributed over the gap and 'bulk' frequency ranges, which is consistent with the fact that the corners states based on the trimers are mixed with the 'bulk' states based on tetramers. Note that, the chiral symmetry of the system is broken due to the introduced connecting tubes [see SI, section 12], which results in the observation that the most intensity of the edge response (blue curve) is distributed on one side of the red curve.

To further demonstrate the HOTIs in the fractal model, we measure the field distributions of the outer corner, inner corner, edge and 'bulk' states and plot the results in Fig. 3(c–f). We can clearly see in Fig. 3(c), the acoustic intensity mainly localizes at four outer corners with the fixed frequency of 5382Hz, which is the significant feature of the traditional HOTIs. Then, sightly off the peak of outer corner response, we find that the sonic intensity at the frequency of 5258Hz is distributed at different inner corners as shown in Fig. 3(d), which presents the properties of self-similarity and the fractional dimensions. Note that there is considerable overlap between the field distributions of the outer and inner corner states due to the fact that the eigen-states are within a small range of frequency. In contrast, the acoustic intensities largely locate at the edges (e) and the 'bulk' (f) with the frequencies within the blue and grey curves. Note that the field intensity of the edge states as shown in Fig. 3(e) is segmented, due to limitation of the fabrication of our acoustic lattice [see SI, section 12]. These experimental results agree well with numerical simulations [see SI, section 14].

In conclusion, we have experimentally demonstrated a higher-order topological insulator in an acoustic fractal lattice. Different from the 2D higher-order topological insulator, the fractal HOTI supports not only 0D outer corner states, but also abundant inner corner states, whose dimension is the same as that of the lattice. Consequently, the codimension in our fractal system is 1.89 for the outer corner states and 0 for the inner corner states, which by definition indicate the fractional-order topological insulator and 'zero'-order topological insulator. The above unconventional results are significantly different from the model with randomly deleting sites in a 2D square lattice [see SI, section 15] and can be further generalized into higher dimensions [see SI, section 16], which unveils that the fractality plays a vital role. This report together with the different topological phases proposed in many types of fractals manifest that the topological fractal insulators are indeed a new family of nontrivial states with the unique features that are distinct from the conventional topological insulators. Moreover, our work provides a feasible platform to investigate the interplay between the fractality and higher-order topological phases and may bring new frontiers in interactions with non-Hermiticity, nonlinearity and quantum optics.

**References:**

[1] W. A. Benalcazar, B. A. Bernevig, and T. L. Hughes, *Quantized Electric Multipole Insulators*, Science **357**, 61 (2017).

[2] J. Langbehn, Y. Peng, L. Trifunovic, F. von Oppen, and P. W. Brouwer, *Reflection-Symmetric Second-Order Topological Insulators and Superconductors*, Physical Review Letters **119**, 246401 (2017).

[3] Z. Song, Z. Fang, and C. Fang, *(D-2)-Dimensional Edge States of Rotation Symmetry Protected Topological States*, Physical Review Letters **119**, 246402 (2017).

[4] F. Schindler, A. M. Cook, M. G. Vergniory, Z. Wang, S. S. P. Parkin, B. Andrei Bernevig, and T. Neupert, *Higher-Order Topological Insulators*, Science Advances **4**, eaat0346 (2018).

[5] M. Ezawa, *Higher-Order Topological Insulators and Semimetals on the Breathing Kagome and Pyrochlore Lattices*, Physical Review Letters **120**, 026801 (2018).

[6] L. Trifunovic and P. W. Brouwer, *Higher-Order Bulk-Boundary Correspondence for Topological Crystalline Phases*, Physical Review X **9**, 011012 (2019).

[7] B. Xie, H.-X. Wang, X. Zhang, P. Zhan, J.-H. Jiang, M. Lu, and Y. Chen, *Higher-Order Band Topology*, Nature Reviews Physics **3**, 520 (2021).

[8] J. Noh, W. A. Benalcazar, S. Huang, M. J. Collins, K. P. Chen, T. L. Hughes, and M. C. Rechtsman, *Topological Protection of Photonic Mid-Gap Defect Modes*, Nature Photonics **12**, 408 (2018).

[9] X. Ni, M. Weiner, A. Alù, and A. B. Khanikaev, *Observation of Higher-Order Topological Acoustic States Protected by Generalized Chiral Symmetry*, Nature Materials **18**, 113 (2019).

[10] X. Zhang, H. X. Wang, Z. K. Lin, Y. Tian, B. Xie, M. H. Lu, Y. F. Chen, and J. H. Jiang, *Second-Order Topology and Multidimensional Topological Transitions in Sonic Crystals*, Nature Physics **15**, 582 (2019).

[11] H. Xue, Y. Yang, F. Gao, Y. Chong, and B. Zhang, *Acoustic Higher-Order Topological Insulator on a Kagome Lattice*, Nature Materials **18**, 108 (2019).

[12] M. Z. Hasan and C. L. Kane, *Colloquium: Topological Insulators*, Reviews of Modern Physics **82**, 3045 (2010).

[13] X.-L. Qi and S.-C. Zhang, *Topological Insulators and Superconductors*, Reviews of Modern Physics **83**, 1057 (2011).

[14] C. W. Peterson, W. A. Benalcazar, T. L. Hughes, and G. Bahl, *A Quantized Microwave Quadrupole Insulator with Topologically Protected Corner States*, Nature **555**, 346 (2018).

[15] X. D. Chen, W. M. Deng, F. L. Shi, F. L. Zhao, M. Chen, and J. W. Dong, *Direct Observation of Corner States in Second-Order Topological Photonic Crystal Slabs*, Physical Review Letters **122**, 233902 (2019).

[16] B. Y. Xie, G. X. Su, H. F. Wang, H. Su, X. P. Shen, P. Zhan, M. H. Lu, Z. L. Wang, and Y. F. Chen, *Visualization of Higher-Order Topological Insulating Phases in Two-Dimensional Dielectric Photonic Crystals*, Physical Review Letters **122**, 233903 (2019).

[17] S. Mittal, V. V. Orre, G. Zhu, M. A. Gorlach, A. Poddubny, and M. Hafezi, *Photonic Quadrupole Topological Phases*, Nature Photonics **13**, 692 (2019).

[18] M. Serra-Garcia, V. Peri, R. Süsstrunk, O. R. Bilal, T. Larsen, L. G. Villanueva, and S. D. Huber, *Observation of a Phononic Quadrupole Topological Insulator*, Nature **555**, 342 (2018).

[19] X. Zhang, B. Y. Xie, H. F. Wang, X. Xu, Y. Tian, J. H. Jiang, M. H. Lu, and Y. F. Chen,

*Dimensional Hierarchy of Higher-Order Topology in Three-Dimensional Sonic Crystals*, Nature Communications **10**, 5331 (2019).

[20] Y. Qi, C. Qiu, M. Xiao, H. He, M. Ke, and Z. Liu, *Acoustic Realization of Quadrupole Topological Insulators*, Physical Review Letters **124**, 206601 (2020).

[21] H. Xue, Y. Ge, H. X. Sun, Q. Wang, D. Jia, Y. J. Guan, S. Q. Yuan, Y. Chong, and B. Zhang, *Observation of an Acoustic Octupole Topological Insulator*, Nature Communications **11**, 2442 (2020).

[22] W. Matthew, N. Xiang, L. Mengyao, A. Andrea, and K. A. B, *Demonstration of a Third-Order Hierarchy of Topological States in a Three-Dimensional Acoustic Metamaterial*, Science Advances **6**, eaay4166 (2022).

[23] S. Imhof et al., *Topolectrical-Circuit Realization of Topological Corner Modes*, Nature Physics **14**, 925 (2018).

[24] J. Bao, D. Zou, W. Zhang, W. He, H. Sun, and X. Zhang, *Topoelectrical Circuit Octupole Insulator with Topologically Protected Corner States*, Phys. Rev. B **100**, 201406 (2019).

[25] S. Liu, S. Ma, Q. Zhang, L. Zhang, C. Yang, O. You, W. Gao, Y. Xiang, T. J. Cui, and S. Zhang, *Octupole Corner State in a Three-Dimensional Topological Circuit*, Light: Science & Applications **9**, 145 (2020).

[26] X. W. Luo and C. Zhang, *Higher-Order Topological Corner States Induced by Gain and Loss*, Physical Review Letters **123**, 073601 (2019).

[27] H. Gao, H. Xue, Z. Gu, T. Liu, J. Zhu, and B. Zhang, *Non-Hermitian Route to Higher-Order Topology in an Acoustic Crystal*, Nature Communications **12**, 1888 (2021).

[28] R. Chen, C. Z. Chen, J. H. Gao, B. Zhou, and D. H. Xu, *Higher-Order Topological Insulators in Quasicrystals*, Physical Review Letters **124**, 036803 (2020).

[29] W. Zhang, D. Zou, Q. Pei, W. He, J. Bao, H. Sun, and X. Zhang, *Experimental Observation of Higher-Order Topological Anderson Insulators*, Physical Review Letters **126**, 146802 (2021).

[30] F. Zangeneh-Nejad and R. Fleury, *Nonlinear Second-Order Topological Insulators*, Physical Review Letters **123**, 053902 (2019).

[31] M. S. Kirsch, Y. Zhang, M. Kremer, L. J. Maczewsky, S. K. Ivanov, Y. v. Kartashov, L. Torner, D. Bauer, A. Szameit, and M. Heinrich, *Nonlinear Second-Order Photonic Topological Insulators*, Nature Physics **17**, 995 (2021).

[32] Z. G. Song, Y. Y. Zhang, and S. S. Li, *The Topological Insulator in a Fractal Space*, Applied Physics Letters **104**, 233106 (2014).

[33] A. A. Iliasov, M. I. Katsnelson, and S. Yuan, *Hall Conductivity of a Sierpiński Carpet*, Physical Review B **101**, 045413 (2020).

[34] M. Fremling, M. van Hooft, C. M. Smith, and L. Fritz, *Existence of Robust Edge Currents in Sierpiński Fractals*, Physical Review Research **2**, 013044 (2020).

[35] Z. Yang, E. Lustig, Y. Lumer, and M. Segev, *Photonic Floquet Topological Insulators in a Fractal Lattice*, Light: Science & Applications **9**, 128 (2020).

[36] M. Benoit, *How Long Is the Coast of Britain? Statistical Self-Similarity and Fractional Dimension*, Science **156**, 636 (1967).

[37] C. Liu et al., *Sierpiński Structure and Electronic Topology in Bi Thin Films on InSb(111)B Surfaces*, Physical Review Letters **126**, 176102 (2021).

[38] *To Be Published*.

[39] S. Pai and A. Prem, *Topological States on Fractal Lattices*, Physical Review B **100**, 155135 (2019).

[40] S. Manna, S. Nandy, and B. Roy, *Higher-Order Topological Phases on Quantum Fractals*, arXiv:2109.03231 (2021).

[41] W. A. Wheeler, L. K. Wagner, and T. L. Hughes, *Many-Body Electric Multipole Operators in Extended Systems*, Physical Review B **10**, 245135 (2019).

[42] B. Kang, K. Shiozaki, and G. Y. Cho, *Many-Body Order Parameters for Multipoles in Solids*, Physical Review B **10**, 245134 (2019).

[43] T. Li, P. Zhu, W. A. Benalcazar, and T. L. Hughes, *Fractional Disclination Charge in Two-Dimensional Cn -Symmetric Topological Crystalline Insulators*, Physical Review B **101**, 115115 (2020).

[44] C. W. Peterson, T. Li, W. Jiang, T. L. Hughes, and G. Bahl, *Trapped Fractional Charges at Bulk Defects in Topological Insulators*, Nature **589**, 376 (2021).

[45] Y. Liu, S. Leung, F.-F. Li, Z.-K. Lin, X. Tao, Y. Poo, and J.-H. Jiang, *Bulk–Disclination Correspondence in Topological Crystalline Insulators*, Nature **589**, 381 (2021).

[46] H. Xue, Z. Wang, Y.-X. Huang, Z. Cheng, L. Yu, Y. X. Foo, Y. X. Zhao, S. A. Yang, and B. Zhang, *Projectively Enriched Symmetry and Topology in Acoustic Crystals*, Physical Review Letters **128**, 116802 (2022).

[47] T. Li, J. Du, Q. Zhang, Y. Li, X. Fan, F. Zhang, and C. Qiu, *Acoustic Möbius Insulators from Projective Symmetry*, Physical Review Letters **128**, 116803 (2022).

[48] Mandelbrot, Benoit, and B, *The Fractal Geometry of Nature*, (WH freeman and Company, San Francisco, 1982).

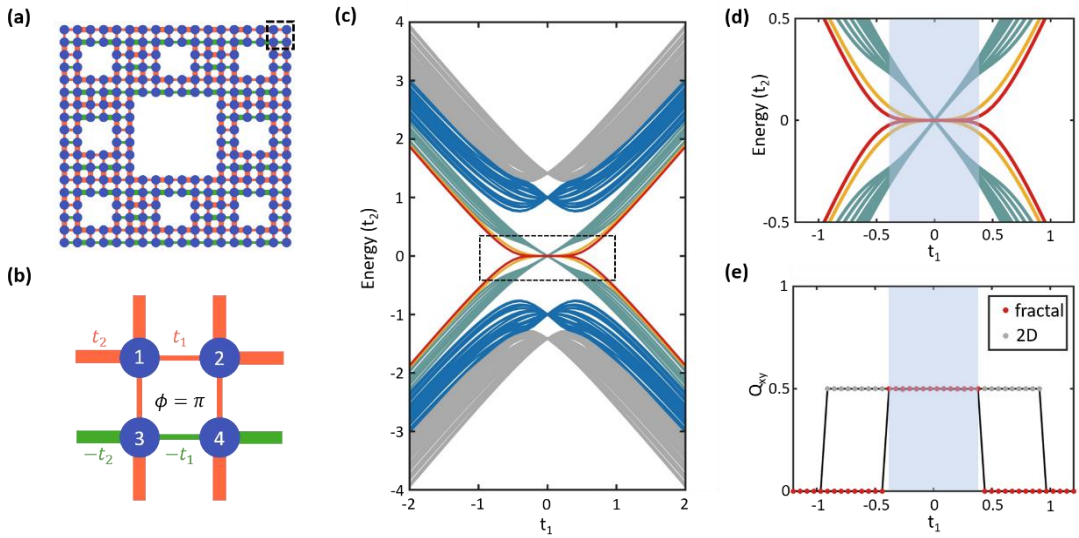

Figure 1. Topological phase diagram of the fractal lattice. (a) Schematic of the fractal Sierpinski carpet. The thin (thick) line represents a hopping with the strength of intra- (inter-) hopping $t_1$ ($t_2$), while the red (green) line represents a positive (negative) hopping. An enlarged view of the dashed box is shown in panel (b). (c) Energy spectrum as a function of intra-hopping $t_1$. Gray and blue curves indicate the bulk' and edge states, while red, yellow, and green curves indicate the outer corner states, inner corner states A and B. (d) Enlarged view of the fractal 'butterfly' in the dashed box shown in panel (c). (e) Squeezed topological region. The gray and red dots correspond to the square and fractal lattice respectively.

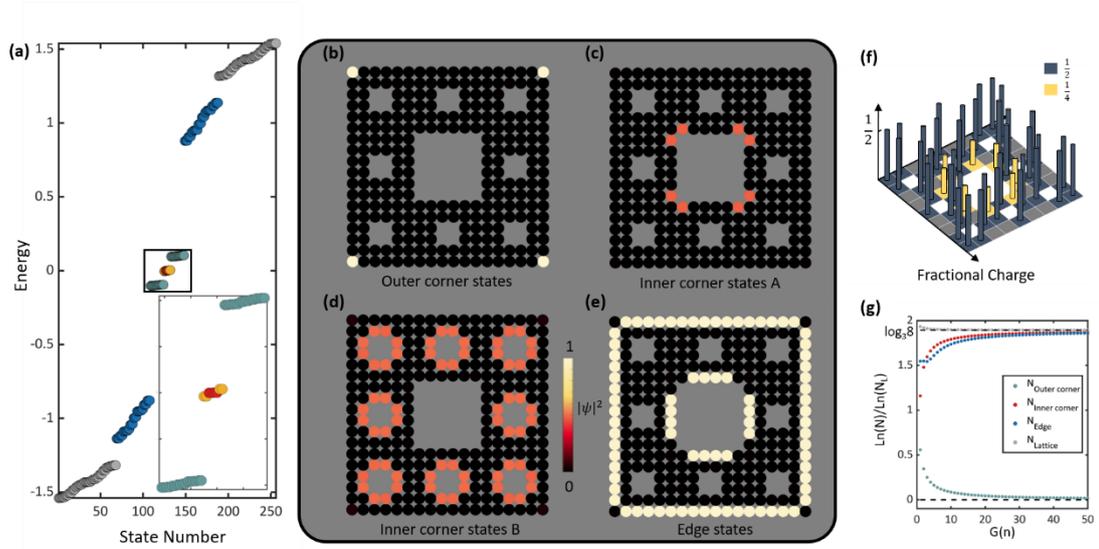

Figure 2. Topological states in the fractal lattice. (a) Energy spectrum with fixed $t_1/t_2 = 0.13$. Inset: an enlarged view of the corner states. (b-e) Sum spatial distribution of the outer corner states, inner corner states A, B, and edge states. (f) Fractional charge in the fractal lattice. There are non-zero fractional charges at the inner and outer corners. (g) Dimensions of the outer corner, inner corner, edge, and fractal lattice as a function of the generation number.

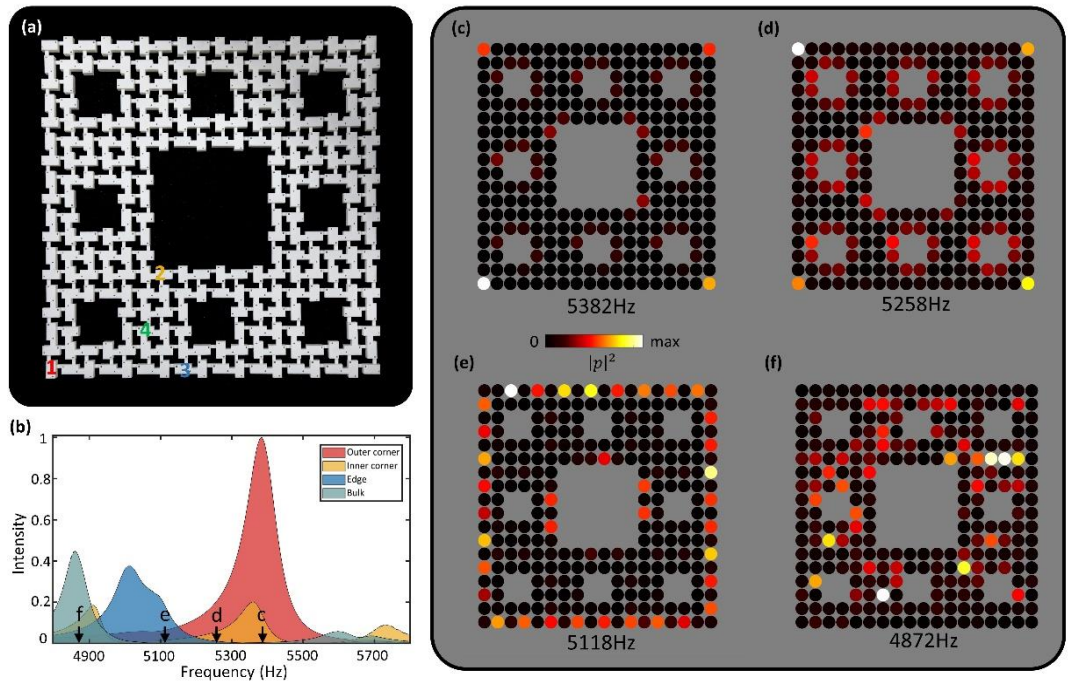

Figure. 3. Experimental observation of the HOTI in an acoustic fractal lattice. (a) Photo of the experimental sample. (b) Measured local density of states at the outer corner, inner corner, edge and 'bulk' sites [marked by '1', '2', '3' and '4' in (a)], which are colored by red, yellow, blue, and green. (c-f) Intensity distributions at the frequencies of 5382, 5258, 5118 and 4872 Hz, which indicate the outer corner, inner corner, edge and 'bulk' states.